\documentclass[11pt,a4paper]{article}
\pdfoutput=1
\usepackage{jheppub}
\usepackage{graphicx}
\usepackage{lipsum}
\usepackage[all]{hypcap}
\usepackage{soul}
\allowdisplaybreaks

\newcommand{\dis}[1]{\begin{equation}\begin{split}#1\end{split}\end{equation}}
\newcommand{\be}{\begin{equation}}
\newcommand{\ee}{\end{equation}}
\newcommand{\bea}{\begin{eqnarray}}
\newcommand{\eea}{\end{eqnarray}}
\newcommand{\beq}{\begin{equation}}
\newcommand{\eeq}{\end{equation}}
\newcommand{\comment}[1]{}

\def\c{{\hat{c}}}
\def\d{{\rm d}}
\newcommand{\Mp}{M_{\rm P}}

\def\V{\hat{V}}

\def\A{{\cal A}}
\def\M{{\cal M}}

\def\R{{\cal R}}
\def\S{{\cal S}}

\begin{document}


{\raggedleft Preprint LMU-ASC 78/18\\}

\title{Heterotic M-Theory from the Clockwork Perspective}

\affiliation[a]{Department of Physics, Pusan National University, Busan 46241, Korea}
\affiliation[b]{Bethe Center for Theoretical Physics and Physikalisches Institut 
der Universit\"at Bonn,\\Nussallee 12, 53115 Bonn, Germany}
\affiliation[c]{Arnold Sommerfeld Center for Theoretical Physics,
Ludwig-Maximilians-Universit\"at M\"unchen, 
\\Theresienstra\ss e 37, 80333 Munich, Germany
}
\affiliation[d]{Institute of Theoretical Physics, Faculty of Physics, University of Warsaw,\\ ul.~Pasteura 5, 02-093 Warsaw, Poland}
 
\author[a]{Sang Hui Im,}
\author[b,c]{Hans Peter Nilles,}
\author[d]{Marek Olechowski}

\emailAdd{imsanghui@pusan.ac.kr}
\emailAdd{nilles@th.physik.uni-bonn.de}
\emailAdd{Marek.Olechowski@fuw.edu.pl}

\abstract{Compactifications of heterotic M-theory are shown to provide solutions
to the weak- and axion-scale hierarchy problems
as a consequence of warped large extra
dimensions. They allow a description that is reminiscent of the so-called 
continuous clockwork mechanism. The models constructed here cover a new
region of clockwork parameter space and exhibit unexplored spectra and 
couplings of Kaluza-Klein modes. Previously discussed models are outside
this region of parameter space and do seem to require an ultraviolet completion
other than that of perturbative higher dimensional $D=10,11$ string- or M-theory. 
A 5D-supergravity description can be given for all explicitly known continuous
clockwork models. The various classes of models can be distinguished through 
the different roles played by vector multiplets and the universal hypermultiplet
in 5D-supergravity.
}

\maketitle

\section{Introduction}

Understanding the origin of small couplings and large hierarchies of scales is a
major challenge in theoretical physics model building. Various mechanisms
have been explored, some based on a slight breakdown of custodial
symmetries (as e.g. supersymmetry),
 others on peculiar properties of extra dimensions. Examples
for the latter include large extra dimensions (LED) \cite{ArkaniHamed:1998rs},
warped extra dimensions (RS) \cite{Randall:1999ee} 
 or the so-called linear dilaton (LD) model 
 \cite{Antoniadis:2001sw}.
A comprehensive discussion  can be formulated in the formalism of the
general continuum clockwork mechanism (GCCW) as described in
Ref. \cite{Choi:2017ncj}.

In the present paper we report on investigations in the framework of
heterotic M-theory \cite{Horava:1996ma}
that makes connection to the GCCW and extends the mechanism in
a nontrivial way that has not yet been explored previously. It also
gives a consistent UV-completion of specific examples of the 
clockwork mechanism in string theory. These main results of our
work will be explained in sections \ref{sec_HW} through \ref{sec_vectors} of the paper.

Let us first give an introduction to the 
clockwork mechanism and the GCCW. The clockwork scheme
can be viewed as a generalisation of the aligned axion  
mechanism \cite{Kim:2004rp,Kappl:2014lra}
originally proposed in the framework of high scale natural inflation.
Its generalisation to the multi-axion case \cite{Choi:2014rja} 
has an interesting application beyond the inflationary picture for the
scale of the QCD axion \cite{Higaki:2015jag}. 
It is also well suited for the discussion of the so-called 
relaxion mechanism  \cite{Graham:2015cka} 
as discussed in Refs.~\cite{Choi:2015fiu} 
and 
\cite{Kaplan:2015fuy}{\footnote{ The name clockwork was first suggested in Ref.~\cite{Kaplan:2015fuy}
}}.
A multi-axion picture with a large number of axions can be connected
to schemes of deconstructions of extra dimensions along the lines
of Refs.~\cite{ArkaniHamed:2001ca,Hill:2000mu}
with a discrete number of sites: the discrete clockwork (DCW).
The transition to a continuous clockwork mechanism (CCW) was
suggested in Ref.~\cite{Giudice:2016yja}.
Some specific properties of the DCW are lost in the generalisation
to a CCW and this leads to some ambiguities in its definition and
interpretation 
\cite{Craig:2017cda,Giudice:2017suc}. 
A comprehensive description of the general picture (GCCW) is given
in Ref.~\cite{Choi:2017ncj}
on which we base our present discussion. Various applications of
the clockwork mechanism have been given in 
\cite{Saraswat:2016eaz, Fonseca:2016eoo, Kehagias:2016kzt, Im:2017eju,
Farina:2016tgd, Ahmed:2016viu, Hambye:2016qkf, vonGersdorff:2017iym,
Coy:2017yex, Ben-Dayan:2017rvr, Hong:2017tel, Park:2017yrn,
Lee:2017fin, Agrawal:2017eqm, Kim:2017mtc, Agrawal:2017cmd, Jeong:2017gdy, Davidi:2017gir, Ibarra:2017tju, 
Patel:2017pct, Dine:2018glh, Cline:2018ebc, Choi:2018dqr, Long:2018nsl, Marques-Tavares:2018cwm, Bonnefoy:2018ibr,
Kim:2018xsp, Niedermann:2018lhx, Davidi:2018sii, Choi:2018mvk, Agrawal:2018mkd, Goudelis:2018xqi, Alonso:2018bcg,
Park:2018kst, Banerjee:2018grm, Craig:2018yld, Co:2018lka, Agrawal:2018vin}.

In this paper, our main focus will be on a subset of the GCCW known as the
general linear dilaton model (GLD) \cite{Choi:2017ncj}. 
It is described by two continuous parameters and it includes the
well-known cases such as LED, RS and LD, but there are many
more possibilities. The goal of the present paper is two-fold:
first to explore the spectrum of GLD models beyond the examples
known up to now and then provide a consistent 
ultraviolet (UV) completion 
within the framework of string theory, if possible.

In this paper we report on progress in both directions:
\begin{itemize}
\item we have found new solutions in the framework of heterotic M-theory,
\item these differ decisively from previously discussed solutions as they exhibit a
new structure for the Kaluza-Klein (KK) spectrum and couplings,
\item we discuss the role of the universal hypermultiplet of compactified
string theory for  the properties for the UV-completion 
of GLD models and derive a bound
on the parameters valid for the models obtained in the framework of heterotic M-theory,
\item previously discussed models are shown to be  mostly outside this bound and
might face difficulties for a UV-completion in perturbative
higher-dimensional $(D=10,11)$
string theory,
\item we provide a 5D-supergravity description for all known models and
discuss the different roles of vector- and hypermultiplets in 5D. 
\end{itemize}

The paper will be structured as follows. In section \ref{sec_GLD} we 
shall give an introduction to the GLD models. We describe the appearance
of the two basic parameters relevant for the creation of 
hierarchical scales and discuss their phenomenological 
consequences for the spectrum of the Kaluza-Klein modes
and the hierarchies of couplings. We reproduce the results for the
previously explored special cases LED, RS and LD. Only for these
three discrete choices of parameters do we have a bottom-up
construction of GLD models. The remaining part of parameter space
still needs explicit realisations of these yet unexplored KK
spectra and couplings. It remains an open question whether more
examples can be realised in quantum field theory and string theory.

A (partial) answer will be given in sections \ref{sec_HW} and \ref{sec_KK}. Here we consider compactifications of heterotic M-theory that could 
explain the hierarchy 
between the Planck scale and hierarchically smaller scales as e.g.
the electroweak scale or the scale of the invisible axion
(which require special choices of parameters that do not 
coincide with the so-called standard embedding). This allows 
a connection towards
the clockwork mechanism and provides new realisations of GLD 
models with phenomenological properties that differ qualitatively
from the previous constructions (LED, RS and LD). As we have a
consistent UV-completion we can give an explicit discussion in
the framework of supergravity on a 5-dimensional (5D) 
manifold with the 5-th dimension being an interval. We 
stress the crucial role of the 5D universal hypermultiplet. 
Models with additional 5D vector multiplets are presented as well 
in section \ref{sec_vectors}.

In section \ref{sec_no_hyper} we shall discuss the clockwork mechanism 
from the 5D perspective and make connection to previous work
that realises the linear dilaton model (LD) within 5D-supergravity
\cite{Kehagias:2017grx,Antoniadis:2017wyh}.
The other known bottom-up constructions (among them RS)  
can be embedded in this scheme. From the 5D-supergravity perspective
here the 5D vector multiplets play a crucial role, while the 
universal hypermultiplet is removed.  Such a situation is impossible
in heterotic M-theory. It remains an open question whether it can be
embedded in any pertrubative higher dimensional ($D=10,11$) string theory.
Section 
\ref{sec_conclusions} contains our conclusions 
and outlook towards a complete classification of GLD models with a consistent
UV-completion.

\section{General Linear Dilaton (GLD) in a nutshell} 
\label{sec_GLD}

In this section, we will give a brief summary of the continuum clockwork and GLD model.
The scalar clockwork action can be written as \cite{Giudice:2016yja}
\dis{
\int \d^4x \left[\sum_{i=0}^{N} \frac{1}{2} (\partial_\mu \phi_i)^2 + \sum_{i=0}^{N-1} \frac{1}{2} m_i^2 (\phi_{i+1} - q\phi_i)^2\right],
} 
where we allow the clockwork gear mass parameter $m_i$ can depend on the site $i$. 
The continuum limit of the action is obtained by taking 
$N \rightarrow \infty$ and introducing 5-th continuous coordinate, 
$y$, with
\begin{equation}
\begin{array}{c}
\displaystyle
\sum_i \rightarrow \frac{1}{\Delta r} \int_0^{\pi R} \d y,\qquad
m_i \rightarrow \frac{m(y)}{\Delta r}, \qquad q-1 \rightarrow k \cdot \Delta r,
\\
\\
\displaystyle 
\phi_i(x) \rightarrow  \Phi(x,y) \,\Delta r^{1/2}, 
 \qquad 
 \phi_{i+1} -\phi_i \rightarrow  \partial_y \Phi(x,y) \,\Delta r^{3/2},
\end{array}
\end{equation}
where $\Delta r \equiv \pi R/N$ is the lattice spacing with finite $R$. 
Here we introduce a dimensionless function $m(y)$ to parameterize the site-dependent mass parameter $m_i$. 
The resultant continuum clockwork action is
\dis{ \label{ccw0}
\int \d^5 x \left[ \frac{1}{2} (\partial_\mu \Phi)^2 + \frac{1}{2} m^2(y) (\partial_y\Phi - k\Phi)^2  \right].
}
As we redefine the field $\Phi \rightarrow \Phi \,e^{ky}$, the action can be rewritten as
\dis{ \label{ccw}
\int \d^5 x \,e^{2k y} \left[ \frac{1}{2} (\partial_\mu \Phi)^2 + \frac{1}{2} m^2(y) (\partial_y\Phi )^2  \right].
}
This action can be obtained from 5D diffeomorphism invariant lagrangian when the metric is replaced by 
a certain background value:
\dis{
\int \d^5 x \, \sqrt{-g}  \frac{1}{2} g^{\alpha \beta} \partial_\alpha \Phi \partial_\beta \Phi,
}
with
\dis{
\d s^2 = e^{\frac{4}{3} ky} m^{\frac{2}{3}}(y)\Big(\d x^2 + m^{-2}(y) \d y^2\Big).
}
Throughout this section, we will use $\alpha, \beta, \dots = 0,\dots, 3, 5$ while $\mu, \nu, \dots =0,\dots, 3$.
As for the clockwork gear mass function $m(y)$, let us consider a simple exponential profile $m(y) = e^{ p y}$. Then the corresponding background geometry is
\begin{equation}
\d s^2 = e^{\frac{4}{3} ky} e^{\frac{2}{3}py}(\d x^2 + e^{-2 py} \d y^2) 
\equiv e^{2k_1 y} \d x^2 + e^{2k_2 y} \d y^2, \label{bgg}
\end{equation}
where 
\begin{equation}
k = k_1 + \frac{1}{2} k_2
\,,\qquad
p = k_1 - k_2. \label{kp}
\end{equation}
From the continuum clockwork action (\ref{ccw0}), it is clear that 
$k$ is responsible for generating coupling hierarchies, while $p$ 
controls the clockwork gear masses (KK masses) via the relation 
$m(y) = e^{ p y}$.

The above background geometry can be generated by GLD proposed in \cite{Choi:2017ncj}. 
The model is defined as the 5D dilaton-gravity action with the specific form of dilaton potential:
\dis{\label{GLD_einstein}
{\cal S}=  M_5^3 \int \d^5 x \sqrt{-g} \,\Big(
 \frac{1}{2} \, \R_5 \,-\,& \frac{1}{2}\,   \partial_\alpha S \partial^\alpha S -  \Lambda_b\, e^{-(2 \c/\sqrt{3}) S} \\
 &- e^{-(\c/\sqrt{3}) S}\left[\Lambda_0  \,\frac{\delta(y)}{\sqrt{g_{55}}} 
+\Lambda_\pi  \,\frac{\delta(y-\pi R)}{\sqrt{g_{55}}} \right]  \Big),
}
where the 5-th dimension $y$ is compactified on an orbifold 
$S_1/{\mathbb Z}_2$ with the fixed points at $y=\{0, \pi R\}$, $M_5$ is the 5D Planck mass, $(\Lambda_b, \Lambda_0, \Lambda_\pi)$ are constants, and $\c$ is an arbitrary real parameter\footnote{
$\c=\sqrt{3}c$ where $c$ was introduced in \cite{Choi:2017ncj}.
}. 
If the potentials at the fixed points satisfy the following relation
\dis{ \label{b.c.}
-\Lambda_0 = \Lambda_\pi = \pm 6 \sqrt{\frac{2}{3}\left( \frac{\Lambda_b}{\c^2-4}\right)},
}
it can be shown that there exists a 4D Minkowski background solution\footnote{
For a discussion of inflationary  4D solutions 
we refer to \cite{Kehagias:2016kzt, Im:2017eju}.}.
Thus, we are left with two free parameters, $\c$ and $\Lambda_b$, 
in \eqref{GLD_einstein}. The consequent background solution turns out to be just (\ref{bgg}):
\[
\d s^2 = e^{2k_1 y} \d x^2 + e^{2k_2 y} \d y^2,
\]
where $k_1$ and $k_2$ are determined in terms of $\c$ and $\Lambda_b$:
\dis{ \label{k1k2}
k_1 = \pm \sqrt{\frac{2}{3}\left( \frac{\Lambda_b}{\c^2-4}\right)}, \quad k_2 = \c^2 k_1.
}
Furthermore, the dilaton field is shown to have a linear dilaton background,
\dis{
(\c/\sqrt{3})S = k_2 y.
}

\begin{figure}[t]
\vspace{-6mm}
\includegraphics[width=0.96\textwidth]{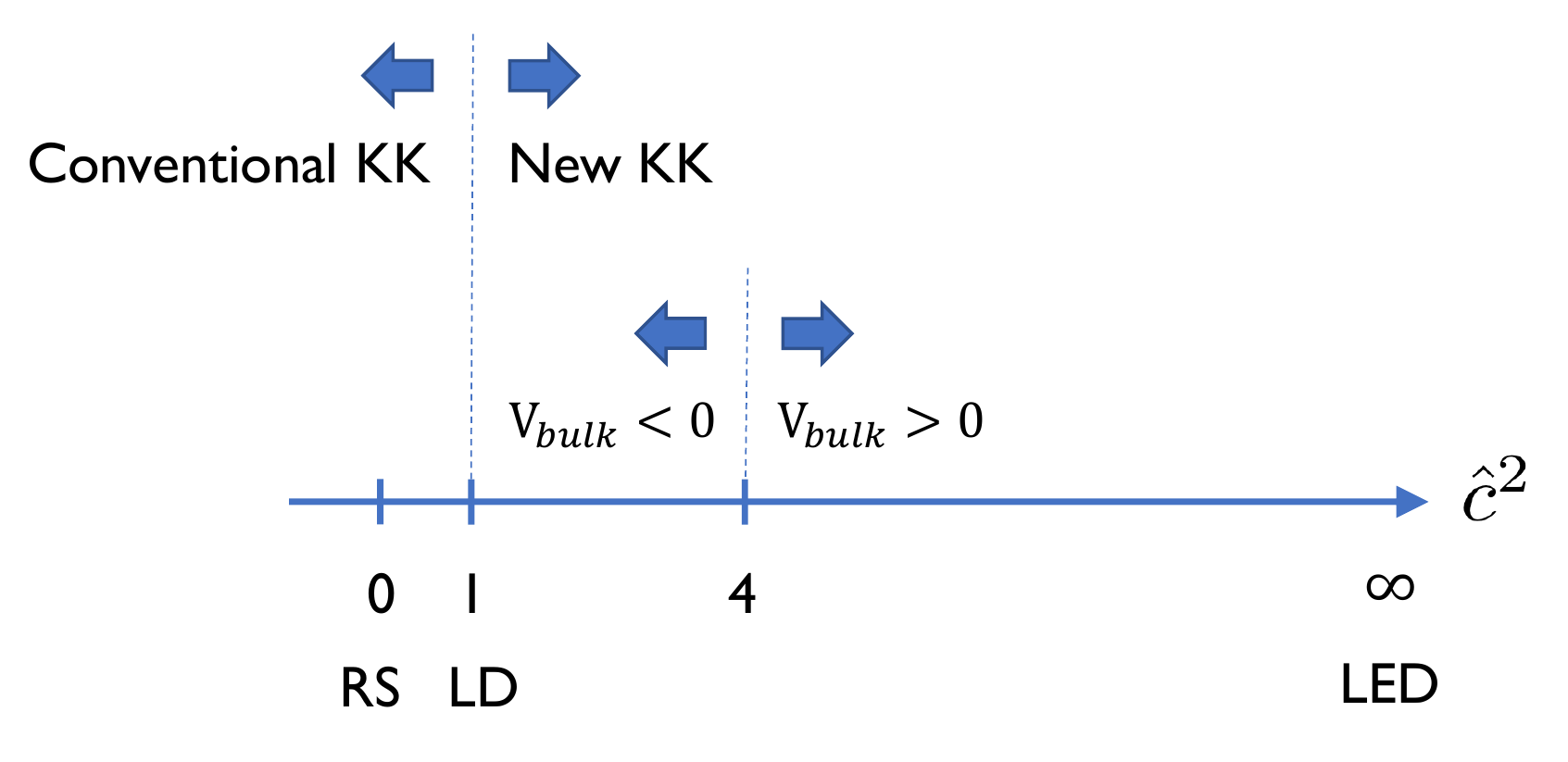}
\vspace{-5mm}
\caption{Critical points of $\c^2$} \label{c_value}
\end{figure}

From the GLD point of view, therefore, we can take $(\c, \Lambda_b)$ as two fundamental parameters to determine
the properties of the clockwork instead of $(k, p)$ or $(k_1, k_2)$.
From the solution (\ref{k1k2}), one can notice that there are critical values for $\c^2$.
For $k_1$ to be real, the bulk potential has to be negative ($\Lambda_b < 0$) for $\c^2 < 4$ while positive ($\Lambda_b >0$) for $\c^2 > 4$. If $\c^2=4$, the dilaton potential should vanish ($\Lambda_b=0$) in order to obtain the 4D Minkowski space, while $k_1$ is not determined by the equation of motion.\footnote{
As we will discuss a specific example in section {\ref{sec_no_hyper}}, 
$k_1$ can be determined by a BPS condition for $\c^2 =4$.
}
 On the other hand, $\c^2=0 \,(k_2=0)$ corresponds to the familiar Randall-Sundrum (RS) model with the AdS$_5$ bulk space.  
 Also, $\c^2=1$ is a special point where $k_1 =k_2$, which corresponds to the linear dilaton model (LD). 
Finally, the flat large extra dimension (LED) can be realized when either the dilaton potential vanishes ($\Lambda_b=0$) or $\c^2 \rightarrow \infty$ so that $k_1 = 0$. 
These critical points are summarized in Fig. \ref{c_value}.

\begin{figure}[t]
\begin{center}
\includegraphics[width=0.96\textwidth]{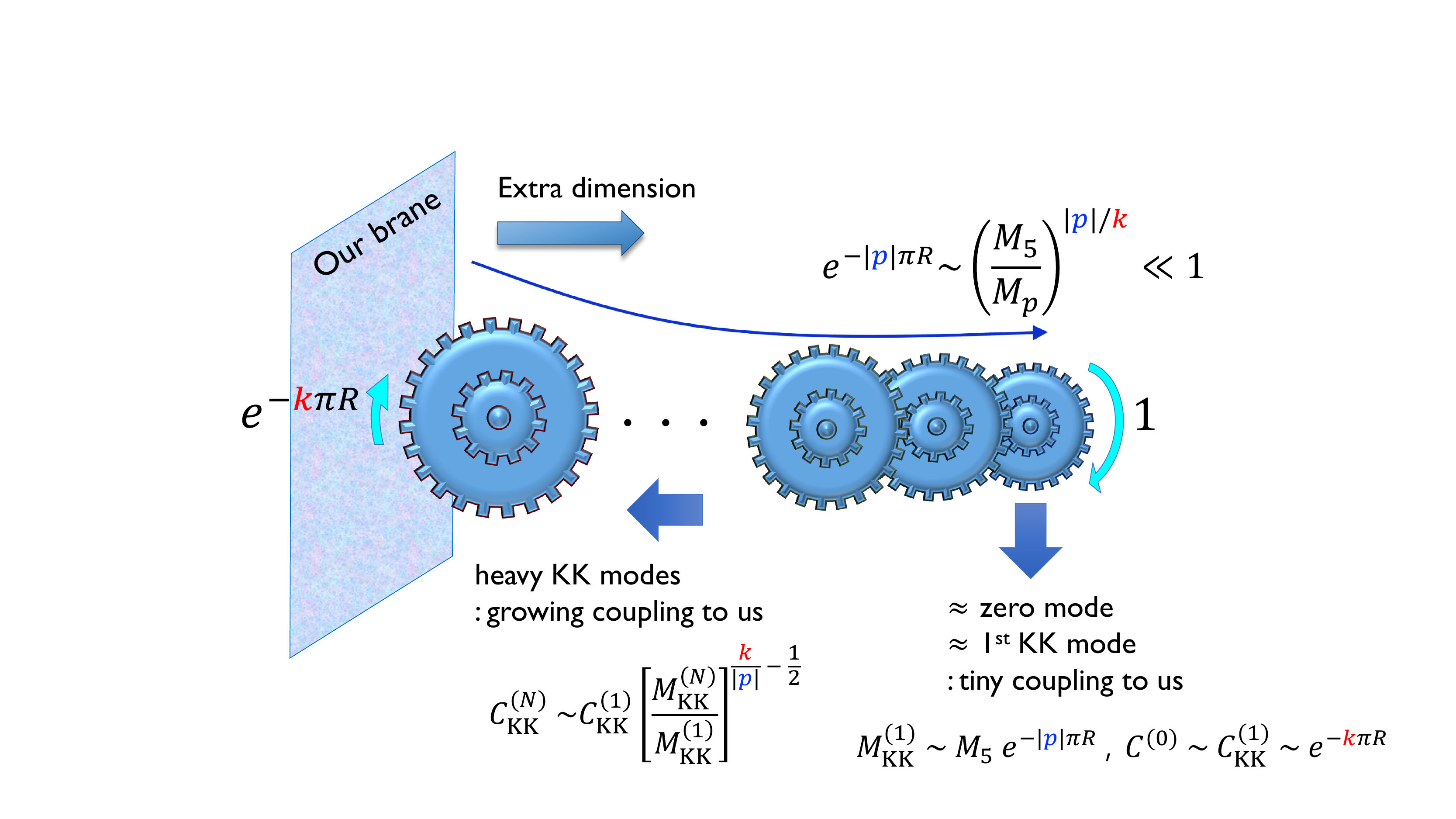}
\vspace{-6mm}
\end{center}
\caption{The KK structure for $\c^2 > 1$} \label{new_KK}
\end{figure}

The point $\c^2=1$ is important in another aspect. In fact, GLD with a finite  $\c^2>1$ predicts a new KK structure which is distinctive from the KK spectra and couplings found in the RS, LD and LED scenarios.\footnote{ Recently, novel collider phenomenology of the LD ($\c^2=1$) scenario compared to RS and LED was extensively studied in \cite{Giudice:2017fmj}. The GLD models with $\c^2>1$ will be even distinctive from the LD scenario.}
A schematic picture for $\c^2 > 1$ is given in Fig.~\ref{new_KK}.
For convention, we will put our brane at $y=0$ and consider the positive $k_1$ solution in (\ref{k1k2}), meaning
a positive $k$ of (\ref{kp}), because this turns out to be able to address the weak scale hierarchy problem. 
The crucial point is that the parameter $p$ of (\ref{kp}) is then negative for $\c^2 >1$. This means that the clockwork gear mass 
function $m(y) = e^{py}$ shows exponentially decreasing profile over the extra dimension. 
Then the zero mode and lightest KK modes are localized near the distant brane $y=\pi R$
so as to have tiny couplings to us. This provides a hierarchy between the 4D Planck mass $\Mp$ and the 5D fundamental scale $M_5$ by $M_5 \sim \Mp e^{-k \pi R}$ with the parameter $k$ of (\ref{kp}). The lightest KK modes' couplings are similarly suppressed. Since the lightest KK modes have much smaller mass scale $(M_{\rm KK}^{(1)} \sim M_5 e^{-|p| \pi R})$ compared to the 5D cutoff $M_5$, the model predicts the KK structure qualitatively similar to LED {\it concerning the lightest modes}. On the other hand, {\it as for heavy KK modes} with masses near the 5D cutoff $M_5$, their couplings are substantially different from LED.
In LED, all KK mode couplings are universally suppressed regardless of their masses. 
In GLD with a finite $\c^2 > 1$, however, heavier KK modes have larger couplings to us proportional to a certain power of their masses.  
This is because the heavy clockwork gears are closer to us as depicted in Fig. \ref{new_KK}.
This may entail interesting phenomenological consequences. We will give a more detailed discussion for the KK structure of $\c^2 > 1$ in section \ref{sec_KK}.

In this paper, we are looking for a motivated UV completion of the GLD clockwork. 
The point $\c^2=1$ (LD) is known to be a 5D approximating theory of the gravity dual of the type II Little String Theory \cite{Antoniadis:2011qw}. 
We now want to see whether there are
more ways to generate clockworks from different types
of string theory.
As we will see in the next section, Ho\v rava-Witten theory \cite{Horava:1996ma}
(or heterotic M-theory) provides a direct realization of 
the GLD clockworks, especially for the region $\c^2 > 1$.

\section{Ho\v rava-Witten model (minimal heterotic M-theory)} 
\label{sec_HW}

The strongly coupled $E_8 \times E_8$ heterotic M-theory may be effectively described in terms of 11-D supergravity with the bosonic part of the action given by
\begin{align}
  \S_{11}
  =
  &-\frac{1}{2\kappa^2} \int_{{\cal M}^{11}} \d^{11} x \sqrt{-g}
  \left(-\R +\frac{1}{24} G_{IJKL} G^{IJKL}
  + \frac{\sqrt{2}}{1728} \epsilon^{I_1 I_2 \dots I_{11}}
  C_{I_1 I_2 I_3} G_{I_4 \dots I_7} G_{I_8 \dots I_{11}} \right)
  \nonumber\\
  &-\frac{1}{8\pi \kappa^2} \left(\frac{\kappa}{4\pi}\right)^{2/3}
  \sum_{i=1}^{2}
  \int_{{\cal M}^{10}_{(i)}} {\d}^{10} x \sqrt{-g} 
  \left( {\rm tr} F_{(i)}^2 -\frac{1}{2} {\rm tr} \R^2 \right).
  \label{11Daction}
\end{align}
In order to obtain ${\cal N}=1$ supersymmetry in 4D we compactify this 
theory on a Calabi-Yau (CY) complex 3-manifold $X^6$. We work in the 
``upstairs'' approach in which ${\cal{M}}^{11}$ is a warped product 
manifold $\M^4\times X^6\times S^1$ where $\M^4$ is the 4D Minkowski 
space-time and the circle $S^1$ is parameterized by the 11-th coordinate $x^{11}$.
If one chooses $x^{11}$ to be in the range $(-\pi r_{11},\pi r_{11}]$
then all
the fields must be symmetric or anti-symmetric and the Lagrangian must
be invariant under parity transformation $x^{11}\to-x^{11}$. 
In the rest of the paper we will use the following conventions 
for the space-time indices:\footnote{
Observe the difference with respect to section 2: the fifth coordinate 
of the 5D (sub)space was denoted by $x^5$ or $y$ and now is denoted by $x^{11}$}
$I,J,\ldots=0,\ldots,9, 11$; $A,B,\ldots=4,\ldots,9$ are tangent to $X^6$;
$\mu,\nu,\ldots=0,1, 2, 3$ are tangent to $\M^4$; 
$\alpha,\beta,\ldots=0, 1,2,3,11$.

Two $E_8$ gauge groups with the field strengths $F_{(i)}$ are localized at
two 10D branes ${\cal M}^{10}_{(i)}$ at $x^{11}_{(1)}=0$ and $x^{11}_{(2)}=\pi r_{11}$,
respectively. By convention we choose the brane at $x^{11}=0$ to support
the gauge sector containing that of the Standard Model.
In order for this theory to be supersymmetric and anomaly-free 
the Bianchi identity for the field strength $G$ is modified such that 
\begin{equation} \label{Bianchi}
{\d}G
=
-\frac{1}{2\sqrt{2}\pi} \left(\frac{\kappa}{4\pi}\right)^{2/3}
\sum_i \left({\rm tr} F_{(i)}^2-\frac12 {\rm tr} \R^2 \right)
\delta\left(x^{11}-x^{11}_{(i)}\right)\,,
\end{equation}
in the leading order of the expansion in $\kappa^{2/3}$,
which results in the following non-zero values of $G$:
\begin{equation}
  \label{Gvalue}
  G_{ABCD} = - \frac{\mu}{48} {\epsilon_{ABCD}}^{EF} \omega_{EF} 
\end{equation}
where $\omega_{EF}$ is the K\"ahler form on $X^6$ while $\mu$ is given by
\begin{equation}
  \mu
  \equiv
  \frac{\sqrt{2}}{\pi V_0} \left(\frac{\kappa}{4\pi}\right)^{2/3}
  \int_{X^6} \omega \wedge \left({\rm tr} F_{(1)} \wedge F_{(1)} -\frac12 {\rm tr} \R \wedge \R \right)
\end{equation}
and $V_0 \equiv \int_{X} {\d}^6 x \sqrt{{\det} g_{AB}}$
is the zeroth order CY volume. Notice that $\mu$ is negative for
the standard embedding of the spin connection in the gauge group
(i.e.~when ${\rm tr}F^2_{(1)}={\rm tr}\R^2$ locally at the brane at $x^{11}=0$)
but may be positive for some non-standard ones. A discussion of the possibilities
can be found in \cite{Stieberger:1998yi}.

The metric on ${\cal{M}}^{11}$ has the form
\begin{equation}
  {\d}s_{11}^2
  =
  (1+\hat{b}) \eta_{\mu \nu} {\d}x^\mu {\d}x^\nu
  + (g_{AB} + \hat{h}_{AB}) {\d}x^{A} {\d}x^{B}
  + (1+\hat{\gamma}) ({\d}x^{11})^2\,,
\end{equation}
where the corrections $\hat{b}$, $\hat{h}_{AB}$ and $\hat{\gamma}$
are functions of only $x^{11}$. The CY part simplifies to
$\hat{h}_{AB}=\hat{h}\,g_{AB}$ if only the universal modulus is taken into
account. Then, in the leading nontrivial order of the $\kappa^{2/3}$ expansion,
the above corrections were found to
be\footnote{ Of course $\hat{\gamma}$ may be changed by a reparametrization
  of $x^{11}$}
\cite{Witten:1996mz,Nilles:1997cm, Nilles:1998sx, Lukas:1997fg, Lukas:1998yy}
\begin{equation}
  \label{bhgamma}
  \hat{b} = \hat{b}_0 \mu |x^{11}|\,,
  \qquad
  \hat{h} = \hat{h}_0 \mu |x^{11}|\,,
  \qquad
  \hat{\gamma} = \hat{\gamma}_0 \mu |x^{11}|\,,
\end{equation}
with
\begin{equation}
  \hat{b}_0 = - \hat{h}_0 = -\frac{\hat{\gamma}_0}{2}=-\frac{\sqrt{2}}{24}\,.
\end{equation}

The effective  4D Planck mass is given by
\begin{align}
  \Mp^2 &= 2 M_{11}^9 \int_0^{\pi r_{11}} {\d}x^{11} \int_{X^6} {\d}^6 x \,
  (1+\hat{b}) \sqrt{{\det}(g_{AB} + \hat{h}_{AB})}  \sqrt{1+\hat{\gamma}}
  \nonumber\\
  &\simeq
  2 M_{11}^9 V_0 \int_0^{\pi r_{11}} {\d}x^{11} \left(1+\frac{1}{2}\hat{\gamma}
  + \hat{b} + 3\hat{h} \right)
  \simeq 2 M_{11}^9 V_0 \pi R_{11}\left(1 + \frac12 \mu \pi R_{11}
  \left|\hat{b}_0 + 3\hat{h}_0\right|  \right)
  \label{MPl}
\end{align}
where $M_{11} \equiv \kappa^{-2/9}$ is the 11D Planck scale and the integration is performed over the CY space $X^6$ and the
11-th interval $I_{11} \equiv S^1/{\mathbb Z}_2$,
and $\pi R_{11}$ denotes the physical length of $I_{11}$,
\begin{equation}
  \label{R11_L}
  \pi R_{11} = \int_0^{\pi r_{11}} {\d}x^{11} \sqrt{1+\hat{\gamma}}
  \simeq \pi r_{11} \left( 1 + \frac{1}{4} \hat{\gamma}_0 \mu \pi r_{11}\right)\,.
\end{equation}
On the other hand, the 6D CY volume scales with $x^{11}$ as
\begin{equation}
  \label{VR11}
  V(x^{11}) = \int_{X^6}  {\d}^6 x \, \sqrt{{\det} (g_{AB} + h_{AB})}
  = V_0 (1+\hat{h})^3
  \simeq V_0 \left(1+3 \mu\, \hat{h}_0 \left|x^{11}\right|\right)\,.
\end{equation}

For negative $\mu$ (e.g.~for the standard embedding) $V$ decreases with
$\left|x^{11}\right|$ which results in an upper bound on the length of
the 11-th dimension, $\pi R_{11}$.
On the other hand, $V$ increases with $\left|x^{11}\right|$ for positive $\mu$. 
In such a case $R_{11}$ may be quite large and the hierarchy problem
may be addressed. In the large $R_{11}$ limit the effective 4D Planck
mass \eqref{MPl} reads
\begin{equation}
  \label{MPl_limit}
  \Mp^2
  \to
  \frac{\sqrt{2}}{12} \mu M_{11}^9 V_0 (\pi R_{11})^2
  = \left[\frac{\sqrt{2}}{12} \left(\frac{\mu}{M_{11}}\right)
    \left(M_{11}^6 V_0\right)\right]M_{11}^4 (\pi R_{11})^2\,.
\end{equation}
$V_0$ is equal to the
CY volume at the observable brane at $x^{11}=0$ so it is related to the
value of the GUT gauge coupling, namely \cite{Witten:1996mz,Nilles:1998sx}
$V_0M_{11}^6=(4\pi)^{2/3}\alpha_{\rm GUT}^{-1}\approx135$. Assuming 
${\frac{1}{16\pi^2}}\int_{X^6}\omega\wedge 
\left({\rm tr}F^2 {-\frac{1}{2}{\rm tr} \R^2}\right) \simeq aV_0^{1/3}$,
where $a$ is an order 1 constant, 
one may estimate the value of the ratio 
$\mu/M_{11} \simeq 0.5a$. 
Using these relations among the massive parameters in the square
bracket in \eqref{MPl_limit} one gets
\begin{equation}
  \label{MPl_approx}
  \Mp^2
  \approx
  8a M_{11}^2 (M_{11}\pi R_{11})^2\,.
\end{equation}
Large 11-th dimension may be used to address the hierarchy problem. 
For example for 
$\pi R_{11}\lesssim100\,\mu$m 
the fundamental mass scale $M_{11}$ of order 
$1.3\,a^{-1/4}\,$TeV
 is enough to obtain the correct value of  $\Mp$.

Notice that the effective 4D Planck mass scales as
$\Mp^2 \sim M_{11}^4 (\pi R_{11})^2$. Such relation is typical
for models with two flat extra dimensions. In the present case it comes from
seven extra dimensions: one large (and flat), $x^{11}$, and six curved, $x^A$, 
with the 6-volume changing linearly with $x^{11}$.

We consider situation when the 11-th dimension is much larger than
the remaining six compact dimensions. So, an effective 5D description
is adequate in such  a limit. In the reduction to 5D we 
will retain only the CY zero modes and universal moduli while keeping the
non-zero mode \eqref{Gvalue}. The 11D metric may be written in the form
\begin{equation}
  \label{5+6Dmetric} 
        {\d}s_{11}^2 = \hat{V}^{-2/3} g_{\alpha \beta} {\d}x^\alpha {\d}x^\beta
        + \hat{V}^{1/3} g_{AB} {\d}x^{A} {\d}x^{B}\,.
\end{equation}
The dependence of the CY volume on 5D coordinates $x^\alpha$ 
is only through the volume modulus field $\hat{V}\equiv V/V_0$. 
The factor $\hat{V}^{-2/3}$ in front of
$g_{\alpha \beta} {\d}x^\alpha {\d}x^\beta$
is introduced to obtain the 5D Einstein frame after reduction.
In this section we consider the simplest version of the 
heterotic M-theory so in the reduction we use only the following 
components of the 3-form field $C_{IJK}$:
\begin{equation}
  \label{C_i}
  C_{\alpha \beta \gamma},
  \qquad
  C_{\alpha AB} = \frac{1}{6} {\cal A}_{\alpha}\, \omega_{AB},
  \qquad
  C_{ABC} = \frac{1}{6} \xi \, \omega_{ABC},
\end{equation}
where $\omega_{ABC}$ is the harmonic $(3,0)$ form on the CY space.
The 3-form field $C_{\alpha \beta \gamma}$ can be dualized to a scalar $\sigma$. 
The resulting theory is the minimal\footnote{
${\cal N}=1$ SUSY from the 5D point
  of view which may lead to ${\cal N}=2$ in 4D after compactification on $S^1$.}
supergravity in 5D with the universal hypermultiplet
$(\hat{V}, \sigma, \xi, \bar{\xi}, \zeta^i)$ and the gravity multiplet
$(g_{\alpha \beta}, {\cal A}_\alpha, \psi_\alpha^i)$ where $\zeta^i$ and
$\psi_\alpha^i$ are the hypermultiplet fermions and gravitini, respectively,
and $i=1, 2$.

Integrating the 11D action \eqref{11Daction} over the CY space $X^6$
one can obtain the effective 5D action. Such reduction retaining the above
mentioned fields and the gauge fields localized at the branes was performed
in \cite{Lukas:1998yy}.
The relevant part for us of this effective action, describing the
gravity--modulus system reads
\begin{align}
  \S_5
  =
  &-\frac{1}{2\kappa_5^2}\int_{{\cal M}^5}\d^5 x \sqrt{-g}
  \left[-{\cal{R}}_5 + \frac{1}{2}\hat{V}^{-2}
    \partial_\alpha\hat{V}\partial^\alpha \hat{V}
    + \frac{\mu^2}{192}\hat{V}^{-2} \right]
  \nonumber\\
  &-\frac{1}{2\kappa_5^2}
  \left\{\frac{\sqrt{2}}{4}\int_{{\cal M}^4_{(1)}}\d^4 x \sqrt{-g}\, \mu\hat{V}^{-1}
  -\frac{\sqrt{2}}{4}\int_{{\cal M}^4_{(2)}}\d^4 x \sqrt{-g}\,  \mu\hat{V}^{-1}\right\}\,, 
\end{align}
where $\kappa_5^2=\kappa^2/V_0$. 
Notice that the non-zero mode of $G$ \eqref{Gvalue} determines
the interaction terms between the graviton and CY volume modulus.
Therefore, this mode is responsible for characteristic low energy
predictions of heterotic M-theory distinctive from
other higher dimensional supergravity/gravity theories,
found e.g~in \cite{Kehagias:1999ju, Teresi:2018eai}.

We would like to analyze the above action in the context of GLD models 
discussed in \cite{Choi:2017ncj} and summarized in the previous section.
Replacing the volume modulus $\hat{V}$ with scalar $S$ defined by
$\hat{V}\equiv\exp(\sqrt{2}S)$ we may rewrite the above 5D action in the general form \eqref{GLD_einstein}
\begin{align}
  \S_5
  =
  &\frac{1}{\kappa_5^2}\int_{{\cal M}^5}{\d}^5 x \sqrt{-g}
  \left[{\frac12{\cal R}_5} -\frac12 \partial_\alpha S \partial^\alpha S
    - \Lambda_b \,  e^{-(2\c/\sqrt{3})S} \right]
  \nonumber\\
  &-\frac{1}{\kappa_5^2}\sum_{i}\int_{{\cal M}_{(i)}^4}{\d}^4x \sqrt{-g}\,
  \Lambda_{(i)} e^{-(\c/\sqrt{3}) S} \label{hGLD}
\end{align}
with
\begin{equation}
  \c^2=6\,,
  \qquad
  \Lambda_b=\frac{\mu^2}{384} \,,
  \qquad
  \Lambda_{(1)}=-\Lambda_{(2)}=\frac{\mu}{4\sqrt{2}}\,. \label{hGLD_cL}
\end{equation}
Thus the dimensional reduction of 
Ho\v rava-Witten model
directly realizes a GLD model with $\c^2 = 6$, and the bulk potential is positive as required for $\c^2 > 4$ as discussed in section \ref{sec_GLD}. Also remarkably, the boundary dilaton potentials satisfy the relation (\ref{b.c.}) to obtain the 4D Minkowski solution. 
This is due to the modified Bianchi identity (\ref{Bianchi}) for the 11D theory to be locally supersymmetric and anomaly free, which implies that
the boundary potentials induced by non-zero magnetic flux $\mu$ must have equal magnitude and opposite sign.

We can reproduce the leading order result (\ref{MPl}) of the 11D action by the effective 5D GLD action (\ref{hGLD}) as well. 
From the 5D background geometry (\ref{bgg}) with the parameters (\ref{k1k2}) and (\ref{kp}), it comes out  to be
\begin{equation}
\Mp^2 = \frac{1}{\kappa_5^2 k} \left(\exp\left(2k\, \pi r_{11}\right) -1\right)
 = M_{11}^9 V_0 \left( \frac{\sqrt{2}}{12}\mu (\pi R_{11})^2 + 2(\pi R_{11}) \right), \label{MPl_GLD}
\end{equation}
where $\pi r_{11}$ is the 11-th coordinate position of the second brane in the coordinate frame used in (\ref{bgg}), and
it is used that
\dis{
\pi R_{11} = \int_0^{\pi r_{11}} \d x^{11} \hat{V}^{-1/3} e^{k_2 x^{11}}
= \frac{3}{2k_2} \left(\exp\left(\frac{2}{3} k_2 \pi r_{11} \right) -1\right), \label{R11}
}
from (\ref{5+6Dmetric}) in order to relate the coordinate radius $r_{11}$ to the physical radius $R_{11}$.
Thus, the 11D leading order result can be obtained as the exact result of the 5D effective theory as pointed out in
\cite{Lukas:1998yy}.

\section{Masses and couplings of Kaluza-Klein (KK) states}
\label{sec_KK}

As emphasized in section \ref{sec_GLD}, $\c^2 = 6 > 1$ will exhibit a new KK structure. The results of the GLD KK spectrum and couplings analysis 
presented in \cite{Choi:2017ncj} may be applied to the lightest KK excitations.
In terms of the parameters defined in section \ref{sec_GLD}, the $n$-th KK mode mass $M_n$ and coupling $C_n$ to our brane turn out to be approximated as 
\bea
  M_n
  &\approx&
  \left(n-\frac14+\frac{k}{2|p|}\right)
  \,\pi|p|\exp\left(-|p| \pi r_{11}\right), \label{Mn} \\
C_n &\approx& \frac{\sqrt{\pi |p|/k}}{\Gamma(k/|p|)} \left[ \frac{\pi}{2} \left(n-\frac{1}{4} + \frac{k}{2|p|} \right)\right]^{\frac{k}{|p|}-\frac{1}{2}} \exp\left(-k\, \pi r_{11}\right), \label{Cn}
\eea
for $p \lesssim -k$ with a positive $k$, where $n=1, 2, \dots$. We have confirmed that these formulae are valid even for \emph{heavy} KK modes up to $n \sim \exp(|p| \pi r_{11})$.\footnote{ The coupling $C_n$ for an exponentially large $n\, (n \lesssim \exp(|p| \pi r_{11}))$ can be different from the formula (\ref{Cn}) by a factor of order one. But we will not be precise on this factor, since it is not important for our discussion.}
In minimal heterotic M-theory, the parameters $k$ and $p$ are equal to
\bea
k &=& \pm\left(1+\frac{\c^2}{2}\right) \sqrt{\frac{2}{3} \left(\frac{\Lambda_b}{\c^2-4}\right)}  = \frac{\sqrt{2}}{12}\, \mu\, , \label{hm_k} \\
p &=& {\rm sgn}(k) (1-\c^2) \sqrt{\frac{2}{3} \left(\frac{\Lambda_b}{\c^2-4}\right)}   =- \frac{5\sqrt{2}}{48}\,\mu\,, \label{hm_p}
\eea
since $\c^2 =6$ and  $\Lambda_b = \mu^2/384$ from (\ref{hGLD_cL}), and the sign $\pm$ is determined by the sign of $\mu$.
Thus for a positive $\mu$, which is necessary to address the hierarchy problem, the condition $p \lesssim -k$ is satisfied to apply the above approximate formulae for the KK spectrum and couplings.
 As was argued in section \ref{sec_GLD}, the lightest KK masses are exponentially suppressed
compared to the 5D fundamental scale $\kappa_5^{-2/3} \sim M_{11} \sim \mu$ by the factor $\exp(- |p| \pi r_{11})$ with the negative $p$ in (\ref{hm_p}) for $\c^2 > 1$.

Using eq.~(\ref{R11}), one can
express the above KK spectrum in terms of the length of the 11-th
dimension, $\pi R_{11}$.
The result (valid for positive $\mu$) reads
\begin{equation}
  \label{MKK2}
  M_n
  \approx
  \left(n+\frac{3}{20}\right)\frac{5\sqrt{2}\pi}{48}\,\mu
  \left(1+\frac{\sqrt{2}}{12}\,\mu\pi R_{11}\right)^{-5/4}.
\end{equation}
Using the relation \eqref{MPl_GLD} between the Planck mass and $R_{11}$, 
we may rewrite \eqref{MKK2} in the large $R_{11}$ limit 
\bea
  M_n
  &\approx&
  \left(n+\frac{3}{20}\right)\frac{5\sqrt{2}\pi}{48}
  \left(6\sqrt{2}\right)^{5/8} \left(M_{11}^6V_0\right)^{5/8}
  \left(\frac{\mu}{M_{11}}\right)^{3/8}
  M_{11}\left(\frac{M_{11}}{\Mp}\right)^{5/4}   \label{MKK3-1} 
  \nonumber\\
  &\approx& 29\, n\, a^{3/8}\,M_{11}\left(\frac{M_{11}}{\Mp}\right)^{5/4} \,. \label{MKK3-2}
\eea
In the case of $N$ large flat extra dimensions (LED), the KK masses scale as
\begin{equation} \label{MLED}
  M_{{\rm LED}, \,\{n_i\} }^2=\left(\sum_i n_i^2\right)
  M_*^2\left(\frac{M_*}{\Mp}\right)^{2/N}\,,
\end{equation}
where $\{n_i\}$ is the set of $N$ numbers describing excitations along
each dimension and $M_*$ is the fundamental mass scale of the model.
Comparing the above formula with the spectrum \eqref{MKK3-2} for large
$n$, we see that asymptotically the KK spectrum of 
the heterotic M-theory is in this case similar\footnote{
  Of course, similar are mass levels but not their multiplicities --
  all levels in the considered M-theory model are non-degenerate.
  }
to that of $N=8/5=1.6$ large extra dimensions.

The GLD clockwork from the simplest version of the heterotic M-theory 
thus resembles  fractional large extra dimensions in terms of 
the KK spectrum, while the hierarchy problem is solved as in the case 
of two large flat extra dimensions. 
Moreover, the KK coupling is predicted to be quite different from LED. 
To see this, let us express (\ref{Cn}) in terms of the KK mass (\ref{Mn}):
\begin{equation}
C_n \approx \frac{\sqrt{\pi |p|/k}}{\Gamma(k/|p|)} \left(\frac{M_n}{2|p|} \right)^{\frac{k}{|p|}-\frac12} e^{-\frac{1}{2} |p| \pi r_{11}}
\approx 97\,a^{-1/2}\, \left( \frac{M_n}{M_1}\right)^{3/10} \left(\frac{M_{11}}{\Mp} \right)\, , 
\label{Cn2}
\end{equation}
where $M_1$ is the first KK mass.
So the coupling (\ref{Cn2}) is similar to the LED KK coupling $C_{{\rm LED},\, \{n_i\}}= M_*/\Mp$ for
the light KK states with $M_n \sim M_1$. But it grows for heavier  KK states.  
 For instance, if we consider a KK state whose mass is near the fundamental cut-off $M_n \sim M_{11}$, the coupling becomes
\dis{
C_n (M_n \simeq M_{11}) 
\approx 
34\,a^{-49/80}\, 
\left(\frac{M_{11}}{\Mp}\right)^{5/8} \gg \,\frac{M_{11}}{\Mp}\, .
}

The phenomenological implications of this novel KK-structure have to be studied further. 
If the scale $M_{11}$ is small enough this might lead to the production of
KK-excitations of the graviton at high energy colliders like the Large Hadron
Collider (LHC) at CERN.
To estimate this production, we consider
 the branching ratio for emitting a KK-graviton of mass scale $M_n$
in a physical process 
with available energy $E$. This can be estimated as
\dis{
{\rm Br}(M_n) \sim E^2 \left(\frac{C_n}{M_{11}} \right)^2 \left(\frac{M_n}{\Delta M_n}\right)
\sim \left(\frac{E}{M_{11}}\right)^2 \left(\frac{M_n}{M_{11}} 
\right)^{8/5},
}
where $\Delta M_n$ is the mass gap of the KK-states and the factor $(M_n/\Delta M_n)$ accounts for the multiplicity of the KK states. Thus  for a collison energy $E$ comparable to the string scale $M_{11}$, KK-gravitons of mass $\sim M_{11}$ may be produced at colliders and the 
spectrum can be analysed.
As we pointed out above, the string scale $M_{11}$ can be as low as a few TeV for the
models under consideration.

\section{Heterotic M-theory with vector multiplets} 
\label{sec_vectors}

In the previous two sections, we have shown that the 5D effective 
theory of heterotic M-theory in its simplest form realizes
the GLD with $\c^2 = 6$. The model was obtained based on the CY volume modulus and graviton interactions. 
Now we want to consider a role of more K\"ahler moduli besides the CY volume modulus i.e.~we consider models compactified on CY space with the Hodge 
number $h_{(1,1)}>1$. 
We will show that in the simplest cases, two more solutions $\c^2 = 7, 10$ are obtained. 

For a Calabi-Yau three-fold $X^6$ with the Hodge number $h_{(1,1)}$, the internal metric $g_{a\bar{b}}$ and K\"{a}hler form $\omega_{a \bar{b}}= i g_{a \bar{b}}$ ($a, b, \dots$ and $\bar{a}, \bar{b}, \dots$ are holomorphic and anti-holomorphic indices on the CY space, respectively) can be expanded with a basis $\omega_i \in H^{(1,1)}(X^6) \,\,(i=1, \dots, h_{(1,1)})$,
\dis{
\omega = t^{i} \omega_i\,,
}
where $t^{i}$ are the K\"{a}hler moduli. 
Then the internal CY space volume $V$ is
\dis{
V = \int_{X^6} \d^6 x \sqrt{\det g_{AB} } = \frac{1}{3!}\int_{X^6} \omega \wedge \omega \wedge \omega = \frac{1}{6} V_0 \, d_{i j k} t^i t^j t^k\,,
}   
with the CY intersection numbers
\dis{
d_{ijk} \equiv \frac{1}{V_0} \int_{X^6} \omega_i \wedge \omega_j \wedge \omega_k\,.
}
If we rescale $t^i = \hat{V}^{1/3} X^i$, we can separate the volume modulus $\hat{V} (\equiv V/V_0)$ 
from the other K\"ahler moduli $X^i$ which satisfy the constraint
 \dis{ \label{vctr}
1 = \frac{1}{6} d_{ijk} X^i X^j X^k\,,
}
so that there are $h_{(1,1)}-1$ independent K\"ahler moduli apart from the volume modulus.

The 3-form field component $C_{\alpha A B}$ in (\ref{C_i}) can be also expanded as
\dis{
C_{\alpha AB} = \frac{1}{6} {\cal A}_{\alpha}^i \, \omega_{i AB} \, .
}
Then in the dimensional reduction to the 5D effective theory, one of the 5D vectors $\A_\alpha^i \,(i=1, \dots, h_{(1,1)})$ is identified as the graviphoton in the 5D gravity multiplet, and the remaining $(h_{(1,1)}-1)$ 5D vectors compose $(h_{(1,1)}-1)$ 5D vector multiplets with the K\"ahler moduli $X^i$.  
Thus in the resulting 5D theory, we have vector multiplets in addition to the universal hypermultiplet and the gravity multiplet considered in the previous section.

In fact, we can also have $h_{(2,1)}$ 5D hypermultiplets constructed out of $h_{(2,1)}$ complex structure moduli $g_{ab}$ and 3-form field components $C_{a b \bar{c}}$.
However, it turns out that these hypermultiplets do not contribute to the scalar potential,
so they are irrelevant for our discussion.

The reduction of the 11D action (\ref{11Daction}) to the 5D effective action retaining the additional vector multiplets was elaborated in \cite{Lukas:1998tt}. The part relevant for us describes the following 
gravity-K\"ahler moduli system:
\dis{ \label{5Daction_vector}
\S_5 =  &-\frac{1}{2\kappa_5^2}\int_{{\cal M}^5}\sqrt{-g}\left[-\R_5+ \frac{1}{2}\V^{-2}\partial_\alpha \V\partial^\alpha \V
+G_{ij}(X) \partial_\alpha X^i \partial^\alpha X^j+ \frac{1}{128}\V^{-2}G^{ij}(X) \mu_i \mu_j \right] \\
&-\frac{1}{2\kappa_5^2}\left\{\frac{\sqrt{2}}{4}\int_{{\cal M}^4_{(1)}}\sqrt{-g}
                   \, \V^{-1}\mu_i X^i-\frac{\sqrt{2}}{4}\int_{{\cal M}^4_{(2)}}\sqrt{-g}\,
                   \V^{-1}\mu_i X^i\right\} \,,
}
where $X^i \,(i=1, \dots, h_{(1,1)})$ are subject to the constraint (\ref{vctr}), and
\begin{equation}
G_{ij} (X) 
= -\frac{\hat{V}^{2/3}}{2} \frac{\partial^2}{\partial t^i \partial t^j} \ln \hat{V}
=-\frac{1}{2} \left[ d_{ijk} X^k - \frac{1}{4} (d_{i lm} X^l X^m) (d_{j n p} X^n X^p)\right] \,,
\end{equation}
and the previous flux parameter $\mu$ is generalized to 
\dis{ \label{mu_i}
\mu_i \equiv \frac{\sqrt{2}}{\pi V_0} \left( \frac{\kappa}{4\pi}\right)^{2/3} \int_X \omega_i \wedge \left({\rm tr} F_{(1)} \wedge F_{(1)} -\frac{1}{2}{\rm tr}\R\wedge \R\right).
}
The important part of the above action is the bulk and boundary scalar potentials arising from non-zero flux parameters $\mu_i$. In particular, the bulk potential derives from non-zero values of the internal components of $G$ as a solution to the modified Bianchi identity (\ref{Bianchi}):
\dis{  \label{Gvalue_2}
G_{ABCD} 
&= -\frac{\hat{V}^{2/3}}{32} G^{ij} \mu_i\, {\epsilon_{ABCD}}^{EF} \,\omega_{jEF} \,\epsilon(x^{11})\, .
}   

As a simple choice, let us first consider $h_{(1,1)}=2$ and the CY intersection number 
$d_{112} \neq 0$ while the other components $d_{ijk}$ vanish. Then the constraint (\ref{vctr}) is
\dis{
d_{112} (X^1)^2 X^2 = 2\,.
}
So we can write
\dis{
X^1 = \frac{1}{\beta} e^{-b S_1},\quad X^2 = \frac{1}{\beta} e^{2b S_1},
}
where $\beta = (d_{112}/2)^{1/3}$, and it turns out that $b=1/\sqrt{3}$ to canonically normalize $S_1$.
On the other hand, we replace the volume modulus $\V$ with scalar $S_V$ defined by $\V=\exp\left(\sqrt{2} S_V\right)$ as before, which corresponds to a canonical normalization for $S_V$.
Then the scalar potential turns out to be
\bea
V_{\rm bulk} &=& \frac{1}{256 \kappa_5^2} \,e^{-2\sqrt{2} S_V}\left[ \,\frac{\mu_1^2}{\beta^2}\, e^{-\frac{2}{\sqrt{3}} S_1} + 2 \,\frac{\mu_2^2}{\beta^2}\, e^{\frac{4}{\sqrt{3}} S_1} \right] \,,\\
V_{\rm boundary} &=& \frac{\sqrt{2}}{8 \kappa_5^2} \, e^{-\sqrt{2} S_V}\left[  \frac{\mu_1}{\beta} e^{-\frac{1}{\sqrt{3}} S_1} + \frac{\mu_2}{\beta}\, e^{\frac{2}{\sqrt{3}} S_1} \right] \left[ \delta(x_{11})-\delta(x_{11}-\pi r_{11}) \right]\,.
\eea

If both $\mu_1$ and $\mu_2$ are non-zero, $S_1$ is stabilized at $\sqrt{3} S_1 = \ln(\mu_1/{2} \mu_2)$. Then we recover the GLD potential (\ref{hGLD}) of the previous section with the GLD dilaton $S$ identified as $S_V$ and $\mu = 3 (\mu_1^2 \mu_2/2d_{112})^{1/3}$.
So in this case, $\c^2=6$ as before.
On the other hand, if one of the flux parameters $\mu_i$ vanishes, one can find a new solution.
If $\mu_1=0$, a new run-away direction $-2\sqrt{2}S_V + 4S_1/\sqrt{3}$ appears in the bulk potential, which is to be identified as the GLD dilaton $S$. In other words, we obtain the GLD action (\ref{GLD_einstein}) 
with 
\bea
 \frac{\c}{\sqrt{3}}S &=& \sqrt{2} S_V - \frac{2}{\sqrt{3}} S_1\, ,  \label{new_S1}\\
\c^2 &=& 3 \left[\left(\sqrt{2}\right)^2 + \left(\frac{2}{\sqrt{3}}\right)^2\right] = 10 \, , \label{new_c1}
\eea
where $\c$ is determined to canonically normalize $S$ as in (\ref{new_c1}). Therefore, we find another GLD with $\c^2=10$ and $\Lambda_b = (2/d_{112})^{2/3} \mu_2^2/128$, while the boundary potential satisfies the condition (\ref{b.c.}) for the 4D Minkowski background. 
Note that the bulk potential is positive as required for $\c^2 > 4$.

Another new solution is obtained when $\mu_2=0$. Similarly to $\mu_1=0$ case, we get the GLD action
(\ref{GLD_einstein}), this time with
\bea
 \frac{\c}{\sqrt{3}}S &=& \sqrt{2} S_V + \frac{1}{\sqrt{3}} S_1\, ,  \label{new_S2}\\
\c^2 &=& 3 \left[\left(\sqrt{2}\right)^2 + \left(\frac{1}{\sqrt{3}}\right)^2\right] = 7 \, .
\eea  
This is a GLD with $\c^2 = 7$ and $\Lambda_b = (2/d_{112})^{2/3} \mu_1^2/256$ 
while satisfying the boundary condition (\ref{b.c.}). 
The bulk potential is positive as well for $\c^2 > 4$.

The above solutions have different scaling for the hierarchy of the mass scales compared to the case
without vector multiplets. Generalizing the formulae (\ref{MPl_GLD}) and (\ref{R11}), it turns out that\footnote{ This formula is strictly valid only 
for ${\hat c}^2 \ge 6$ -- see eq.~\eqref{hm_bound} below.
}
\dis{
\Mp^2 \sim M_{11}^2 \times (\pi R_{11} M_{11})^{(\c^2+2)/(\c^2-2)}.
\label{NeffMPl}
}
Therefore, $\c^2=7, 10$ correspond to 1.8, 1.5 flat extra dimensions, respectively\footnote{ For smaller effective number of such extra dimensions larger 
values of $M_{11}$ are necessary to get the correct value of 4D Planck 
mass. For $\pi R_{11}\sim100\,\mu$m the 11D fundamental scale must be of 
order ${\cal O}(10)\,$TeV for $\c^2=7$ and ${\cal O}(100)\,$TeV for $\c^2=10$.
}. 
Analogously to the previous case, the corresponding KK spectra are matched
to different (fractional) number of extra dimensions.
For large $n$ the spectrum 
is similar to that of $N=3/2=1.5$ large flat extra dimensions for 
$\c^2=7$ and $N=4/3$ for $\c^2=10$. Generally the number of extra dimensions $N$ corresponding to KK spectrum is related to $\c$ by the following formula
\begin{equation}
N=\frac{\c^2+2}{\c^2-1}\,.
\label{NeffKK}
\end{equation}
%
%
One should note that for heterotic M-theory for which $\c^2\ge6$ 
(see eq.~\eqref{hm_bound} and discussion below) both effective numbers of 
dimensions, one related to the hierarchy of scales in eqs.~\eqref{NeffMPl} and another related to the KK spectrum given by \eqref{NeffKK}, are monotonically 
decreasing functions of $\c^2$.

We have therefore found two more GLD clockwork solutions $\c^2 = 7, 10$ by the example of $h_{1,1}=2$ with $d_{112} \neq 0$ and $d_{ijk}=0$ otherwise. By a similar procedure, one can find that the same solutions are obtained for another simple example of $h_{1,1}=3$ and $d_{123} \neq 0$ while $d_{ijk}=0$ otherwise. 
Although we do not prove it rigorously, it seems quite generic that we recover the solution $\c^2=6$ when
all flux parameters $\mu_i$ are non-zero, because the K\"ahler moduli involved in the vector multiplets are stabilized so that only the volume modulus  plays the role of the GLD dilaton. On the other hand, if one of the flux parameters vanishes, a new run-away direction like (\ref{new_S1}) or (\ref{new_S2}) appears in the bulk potential so as to be identified as the GLD dilaton with a new $\c^2$.  
At any rate, the possibility to get more different solutions from higher $h_{1,1}$ or more complicated $d_{ijk}$ is still open.

From the above example, one can observe that there is a lower bound for $\c^2$:
\dis{
\c^2 \geq 3(\sqrt{2})^2 = 6\,. 
\label{hm_bound}
}
This bound is due to the presence of the volume modulus $S_V$. The effect of the other K\"ahler moduli $S_i$ either does not change or increases $\c^2$. 
In fact, this is a generic lower bound for heterotic M-theory, since the general scalar potential in  (\ref{5Daction_vector}) is multiplied by the overall volume modulus factor.  
Therefore, LD ($\c^2=1$) or RS ($\c^2=0$) scenarios cannot be realized in heterotic M-theory. 
Does this mean that they cannot be consistently embedded in string theory? 
The next section is devoted to a discussion of this question.

\section{Can we avoid the heterotic M-theory bound?} 
\label{sec_no_hyper}

In the previous section, we derived the lower bound $\c^2 \geq 6$ 
in the framework of heterotic M-theory.
It was observed that this bound is due to the presence of the CY volume 
modulus. With a compactification on a CY manifold we arrive at a 5D theory 
and we might try to understand the existence of such a bound in terms 
of 5D supergravity as well. Since the CY volume modulus is contained 
in the 5D universal hypermultiplet, this bound will be related
to the presence of that multiplet.  If we want to avoid the bound we have 
to discuss the role and coupling of the universal hypermultiplet in detail.

As pointed out in \cite{Lukas:1998yy, Lukas:1998tt, Behrndt:2000zh}, a non-zero $G$-flux in heterotic M-theory and/or 11D supergravity leads to gauging the resulting 5D effective supergravity along the axion $\sigma$ direction, which is the dual of the 3-form component $C_{\alpha \beta \gamma}$. 
So the bulk scalar potential can be written in terms of a \emph{gauged} 5D supergravity,
\dis{ \label{gauged}
V_{\rm bulk} = -2 G^{ij}(X) {\rm tr}{\cal P}_i {\cal P}_j + 4 X^i X^j {\rm tr} {\cal P}_i {\cal P}_j + \frac{1}{2} X^i X^j h_{uv} k_i^u k_j^v
} 
with the Killing vector $k_i =  (\mu_i/4) \partial_\sigma$ and the Killing prepotential
\dis{
{\cal P}_i = \left( {\begin{array}{cc}
   \frac{i}{32\hat{V}} \mu_i & 0 \\
   0 & -\frac{i}{32\hat{V}} \mu_i \\
  \end{array} } \right)\,,
}
and $h_{uv}$ is the metric for the universal hypermultiplet $q^u = (\hat{V}, \sigma, \xi, \bar{\xi})$ in which $h_{\sigma \sigma}= 1/(4 \hat{V}^2)$. 
The first two terms in the potential are contributions from the gravity 
and vector multiplets, while the last term is from the universal 
hypermultiplet.  
Now  the last term cancels the second term
so that the bulk potential is given by the first term alone. One can see 
that this first term is the same as the bulk potential in (\ref{5Daction_vector}). 
Observe that this term is non-negative.
As shown in Fig.~\ref{c_value} this implies that  
$\c^2$ has to be greater than $4$ if a GLD clockwork solution exists (as 
explained in section \ref{sec_GLD}).

In the framework of 5D supergravity, however, there exists previous work
\cite{Kehagias:2017grx, Antoniadis:2017wyh} that describes clockwork systems with
$\c^2<4$. The models considered there were constructed exclusively 
in the presence of vector multiplets. To make contact with these investigations
we have to take our 5D system described above and remove the universal
hypermultiplet.  From the viewpoint of the underlying higher-dimensional string theory this might be
problematic, as the universal hypermultiplet is an important ingredient in the
compactifications of heterotic M-theory (and 11D M-theory as well).
Let us nonetheless consider the case where the universal hypermultiplet is removed 
from the system.
This corresponds to set the Killing vector $k_i=0$ and the Killing prepotential ${\cal P}_i = {\rm const}$.
Since the third contribution in (\ref{gauged}) vanishes, the second term now contributes to the bulk potential. Because this second term is non-positive, it can make the bulk potential negative so that $\c^2 < 4$ can be realized. 
Therefore, the heterotic lower bound for $\c^2$ can be avoided if the universal hypermultiplet is decoupled. 
However, we should stress that it is impossible to decouple the 
universal hypermultiplet in heterotic M-theory, because
it is constrained by the boundary gauge couplings of $E_8 \times E_8$, e.g. by perturbativity of the gauge couplings. 

As an alternative we instead might consider  M-theory (in contrast to heterotic M-theory)
in the limit of 11D supergravity and hope that such a situation can be realized there
\cite{Kehagias:2017grx}. It is, however, not clear to us, how this can happen. The situation
might just be a realization of the clockwork in 5D supergravity without a meaningful
embedding in 11D M-theory.

Let us nonetheless discuss the phenomenological consequences 
in the absence of the universal hypermultiplet. 
As discussed above, this makes the second term in (\ref{gauged}) contribute to the bulk potential in addition to the first term. If we again consider the example of the previous section $h_{(1,1)} = 2$ and $d_{112} \neq 0$ while the other $d_{ijk} =0$, the bulk potential becomes    
 \bea
V_{\rm bulk} &=& \frac{1}{256\kappa_5^2} \,\hat{V}^{-2} \left[ \,-\frac{\mu_1^2}{\beta^2}\, e^{-\frac{2}{\sqrt{3}} S_1} - 4 \,\frac{\mu_1 \mu_2}{\beta^2}\, e^{\frac{1}{\sqrt{3}} S_1} \right], 
\eea
where  $\hat{V}$ is a constant. Notice that the bulk potential is now negative.  This is the same potential as found by \cite{Kehagias:2017grx, Antoniadis:2017wyh} in the context of the LD clockwork realization through 5D gauged supergravity with vector multiplets. 
As discussed there, if both $\mu_1$ and $\mu_2$ are non-zero, there is an extremum of the potential where $S_1$ is stabilized. This realizes the RS scenario ($\c^2=0$) with 
negative bulk potential. 
On the other hand, if $\mu_2=0$ (and $\mu_1 \neq 0$), $S_1$ becomes a run-away direction which realizes the GLD action (\ref{GLD_einstein}) with
\bea
 S = S_1\,\,\, {\rm and} \,\,\,
\c^2 = 1\,,
\eea
corresponding to the linear dilaton case (LD).
If $\mu_1 =0$, the bulk potential vanishes. Still, a non-zero warping along the 11-th dimension can be obtained
when $\mu_2 \neq 0$  as discussed in  \cite{Kehagias:2017grx}. Interestingly, the resultant background solution turns out to correspond to the GLD clockwork solution with 
\bea
S = S_1\,\,\, {\rm and} \,\,\,
\c^2 = 4\,,\eea 
while $\Lambda_b =0$ and $k_1$ is determined by $\mu_2$. 
This exhausts the solutions obtained  in the case 
$h_{(1,1)} = 3$ and $d_{123} \neq 0$ while the other $d_{ijk} =0$.

We have thus seen that in the absence of  
the universal hypermultiplet in the 5D supergravity theory, 
the heterotic bound $\c^2 \geq 6$ can be avoided.
In the presence of vector multiplets one can construct the solutions
 RS ($\c^2=0$) as well as LD ($\c^2=1$) and a third solution ($\c^2 =4$).
The possibility to obtain other GLD clockwork solutions with higher $h_{(1,1)}$ or more complicated
$d_{ijk}$ is the subject of further investigations.

\section{Conclusions and Outlook}
\label{sec_conclusions}

GLD models provide a 2-parameter class of potential 
solutions to the weak- or axion-scale hierarchy problems.
Depending on the parameters, the models differ in the properties of KK-masses and
couplings. The GLD set-up has been discussed in a bottom-up construction and it
remains to be seen whether there is a consistent UV-completion in the framework
of string theory. 

\begin{figure}[t]
\vspace{-6mm}
\includegraphics[width=0.96\textwidth]{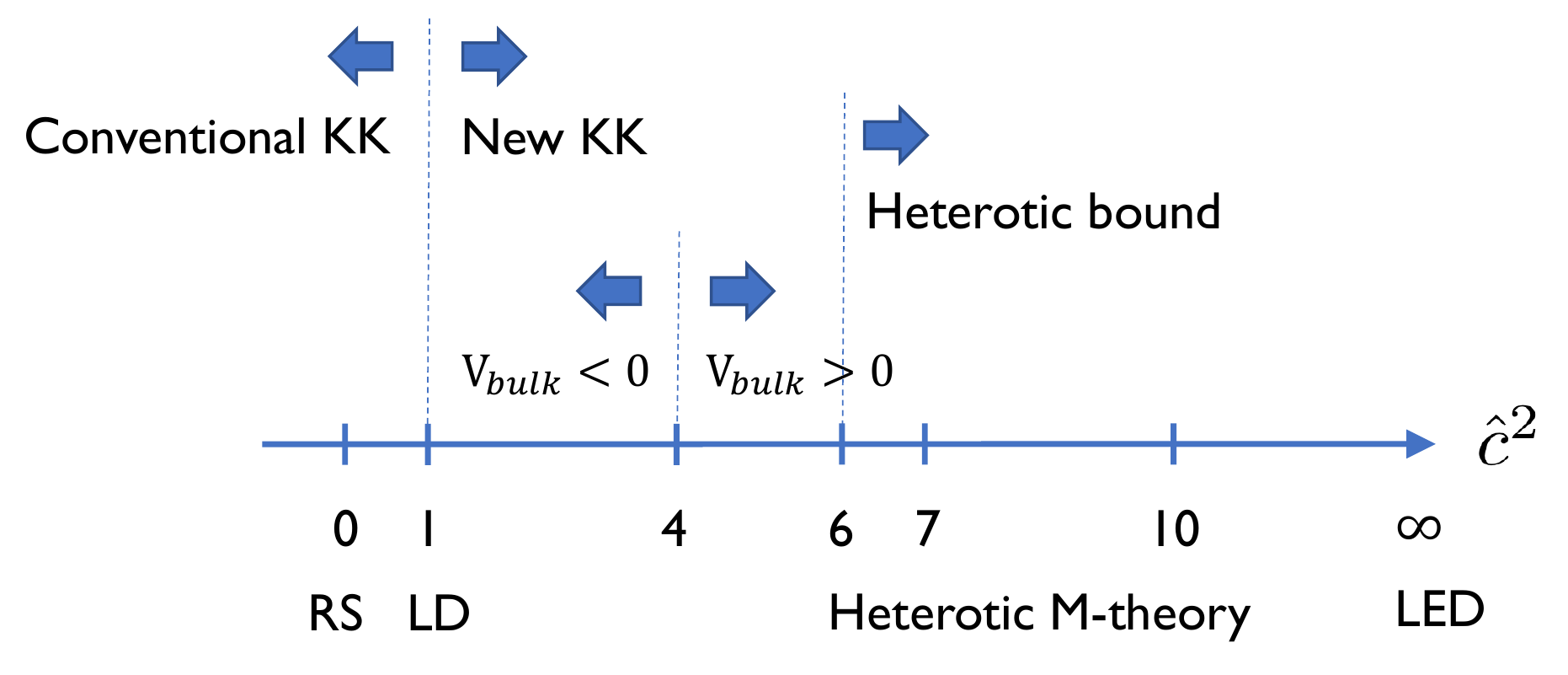}
\vspace{-5mm}
\caption{Critical points of $\c^2$. Heterotic M-theory can realize the GLD clockworks with $\c^2 \geq 6$.} \label{c_value_hm}
\end{figure}

The results of our investigations are summarized in Fig.~\ref{c_value_hm}, which indicate that such a valid UV-completion might only be
possible for some discrete values of $\c^2$. We were able to derive such models in the
framework of compactified heterotic M-theory for the values $\c^2=6, 7, 10$. They show
an unconventional interpretation in the geometrical picture. For the hierarchy of the 
mass scales, the case of $\c^2=6$, for example, appears as a 
model with 2 flat extra dimensions.
However, the corresponding KK-spectrum resembles that of a model with
1.6 extra dimensions.
This unconventional
behaviour strongly influences the couplings of the KK-modes and might have 
interesting applications for the role of axions in heterotic M-theory compactifications
\cite{heterotic_axion}. For consistent models within the framework of 
heterotic M-theory we can derive a lower bound: $\c^2\ge 6$. This bound appears as
a consequence of the presence of the universal hypermultiplet in the 
theory compactified to 5 dimensions.

Previously discussed models, such as RS and LD, violate this bound. In a 5D-supergravity
description they require the presence of vector multiplets (but the absence of the universal
hypermultiplet). It remains to be seen whether this allows a consistent uplift to a
higher dimensional string- or M-theory. 
Models outside the heterotic M-theory bound $\c^2\ge 6$ do apparently require a
different UV-completion. For the linear dilaton model (LD) 
such a completion has been
suggested in the framework of a 6D non-critical and non-local string-like theory
known as "Little String Theory". For a recent discussion and references we refer
the reader to section 2.2 of \cite{Giudice:2017fmj}.
It remains an open problem to understand a possible relation between these
different attempts for a UV-completion.
Other open questions concern the complete classification of the models that 
allow a consistent UV-completion. Up to now we were only able to find solutions
for certain discrete values of $\c^2$.
 
The unconventional properties of the KK-spectrum and the couplings of the
various KK-excitations on the various branes deserve further investigations
which we will discuss in a future publication \cite{heterotic_axion}. 
They might have relevance for 
researches at colliders \cite{Giudice:2017fmj}
as well as axionic couplings within the framework of 
heterotic M-theory in view of a solution of the strong CP-problem.  

\acknowledgments

This work was supported by the German Science Foundation (DFG) within the 
SFB-Transregio TR33 ``The Dark Universe". 
SHI acknowledges support from the National Research Foundation of Korea (NRF) grant 
funded by the Korea government (MSIP) (NRF-2018R1C1B6006061).
SHI also thanks the CERN-Korea TH Institute, 
where valuable comments were given to this work. 
HPN thanks Dieter L\"ust and
the ASC-Center at LMU Munich for hospitality and support.
MO acknowledges partial support from National Science Centre, Poland, 
grants DEC-2015/18/M/ST2/00054 and DEC-2016/23/G/ST2/04301.

%
\newpage

\bibliography{heterotic_CW}
\bibliographystyle{utphys}

\end{document}